\documentclass[pre]{revtex4}
\usepackage{amsmath}
\usepackage{graphicx}
\begin{document}
\title{Reflection of Channel-Guided Solitons at Junctions in  Two-Dimensional Nonlinear Schr\"odinger Equation}
\author{Yusuke Kageyama and Hidetsugu Sakaguchi}
\affiliation{Department of Applied Science for Electronics and Materials,
Interdisciplinary Graduate School of Engineering Sciences, Kyushu
University, Kasuga, Fukuoka 816-8580, Japan}
\begin{abstract}
Solitons confined in channels are studied in the two-dimensional nonlinear Schr\"odinger equation. We study the dynamics of two channel-guided solitons near the junction where two channels are merged.   The two solitons merge into one soliton, when there is no phase shift.  If a phase difference is given to the two solitons, the Josephson oscillation is induced. The Josephson oscillation is amplified near the junction. The two solitons are reflected when the initial velocity is below a critical value.  
\end{abstract}
\maketitle
The one-dimensional nonlinear Schr\"odinger equation has been intensively studied as a typical soliton equation for optical solitons in optical fibers [1] and matter-wave solitons in the Bose-Einstein condensates (BECs)~[2],[3]. 
However,  solitons in two- or three-dimensional nonlinear Schr\"odinger equations were not intensively studied. Two-dimensional solitons exist in guided channels in the two-dimensional nonlinear Schr\"odinger equation.~[4] The guiding channel can be constructed by modifying the profile of the refraction index in an optical planar waveguide.  Cigar-shaped traps are used as a guiding channel to confine matter-wave solitons in BECs. Solitons can propagate along the guiding channel with an arbitrary velocity if the guiding channel is uniform and the norm of the solitons is below a critical value for collapse.  In our previous study~[5], we investigated the reflection of a channel-guided soliton in a tapered channel and the splitting at a branching point where one channel branches into two channels~[6].   
In this study, we investigate the motion of solitons in curved channels and near the junction where two channels merge into a channel. 

The model equation is written as 
\begin{equation}
i\frac{\partial \phi}{\partial t}=-\frac{1}{2}\nabla^2\phi-|\phi|^2\phi+U(x,y)\phi,
\end{equation}
where $U(x,y)$ denotes the potential for confinement, which is used to make a guiding channel. For a straight channel, $U(x,y)=-U_0$ for $-x_0\le x  \le x_0$ and $U(x,y)=0$ for other 
regions. The width of the channel is denoted as $2x_0$ and the depth of the potential is denoted as $U_0$. 
There are stationary soliton solutions $\phi_0(x,y)$ in the straight channel. 
The stationary solution $\phi_0(x,y)$ can be numerically obtained by the imaginary-time evolution of eq.~(1) as was carried out in our previous study~\cite{rf:5}. The two-dimensional soliton can move with any velocity $k_y$ if the initial condition is set to be $\phi_0(x,y-y_0)\exp\{ik_y (y-y_0)\}$. 

In the previous study, we investigated the splitting and reflection of a two-dimensional soliton in a branching channel and designed a branching channel where a soliton can split smoothly into two solitons without changing its velocity. 
If the branching channels merge into a straight channel at a junction point again, a waveguide system, such as the Mach-Zehnder interferometer, can be constructed, as shown in Fig.~1(a). The branching point is $y=-40$ and the junction point is $y=40$ in the channel system in Fig.~1(a). 
A phase shift $\Delta\varphi$ is given to the two split solitons when the center of the solitons goes through the line $y=0$ as $\phi(x,y)\rightarrow \phi(x,y)\exp(i\Delta\varphi)$ for $x<0$ and $\phi(x,y)\rightarrow \phi(x,y)\exp(-i\Delta\varphi)$ for $x>0$. This procedure is an artificial one, which is simple and easy for numerical simulation. In an actual experiment, the phase shift can be effectively given to the two solitons by controlling the profiles of the refraction index and the forms of the two channels similar to the typical Mach-Zehnder interferometer.  When $\Delta\varphi=0$, the two solitons merge again into one soliton at the junction located at $y=40$.  We have performed numerical simulation with the split-step Fourier methods with $128\times 1024$ modes for this rectangular system. 
For nonzero $\Delta \varphi$, a zigzag oscillation of the center of gravity is observed, as shown in Fig.~1(b).   
Figure 1(b) shows the trajectories of the center of gravity of the solitons for $\Delta\varphi=0,\pi/64,\pi/32$ and $\pi/16$ at $k_y=1$. The norm $N$ of the soliton is 5. A zigzag oscillation appears in a region of $y>20$. 
The amplitude of the zigzag oscillation increases with $\Delta\varphi$, but the period of the oscillation is almost the same.  
This zigzag oscillation is interpreted as a remnant of the Josephson oscillation  that appears before the junction point at $y=40$. The Josephson oscillation occurs owing to the tunnel effect through the potential wall of height $U_0$ between the two channels. The  Josephson oscillation is explained in detail in the following.  

When the initial velocity $k_y$ is decreased for a fixed phase shift value, the solitons are reflected before the junction point at $y=40$. Figure 1(c) shows the time evolution of $|\phi|$ along the center lines of the left and right channels at $k_y=0.37$ for $\Delta\varphi=\pi/256$. Before the reflection, the amplitudes of the two solitons in the left and right channels are almost the same, but the difference between the amplitudes of the two solitons increases after the reflection, and the soliton becomes localized in one of the two channels.  The critical velocity $k_{yc}$ is 0.38 for $\Delta\varphi=\pi/256$. The relationship between $k_{yc}$ and the phase shift $\Delta\varphi$ is shown in Fig.~1(d). Note that the horizontal axis is plotted with a logarithmic scale. Figure 1(d) implies that the critical velocity $k_{yc}$ increases rapidly near $\Delta \varphi=0$. 

If there is a phase difference between two solitons located on different channels, the Josephson oscillation occurs~[7],[8], because the two solitons interact with each other by the tunnel effect through the potential barrier of height, $U_0$. 
We can take an ansatz for the form of $\phi$ as $\phi(x,y)=u(y,t)\exp[-\{x-\eta(y)\}^2/(2b^2)]\exp(-i\mu t)+v(y,t)\exp[-\{x+\eta(y)\}^2/(2b^2)]\exp(-i\mu t)$, where $\pm \eta(y)$ is the $x$-coordinate of the central point in the right and left channels at $y$.  The substitution of the form into the Lagrangian (2) and the variational principle: $\partial/\partial t(\delta L/\delta u_t)=\delta L/\delta u$, $\partial/\partial t(\delta L/\delta v_t)=\delta L/\delta v$, yield approximately coupled equations for $u$ and $v$:
\begin{eqnarray}
i\frac{\partial u}{\partial t}&=&-\frac{1}{2}\frac{\partial^2u}{\partial y^2}-c(|u|^2+g|v|^2)u-d(v-u),\nonumber\\
i\frac{\partial v}{\partial t}&=&-\frac{1}{2}\frac{\partial^2v}{\partial y^2}-c(|v|^2+g|u|^2)u-d(u-v),
\end{eqnarray}
where $g=2\exp(-2\eta(y)^2/b^2)$ and $d=2U_0\exp(-2\eta(y)^2/b^2)$. Here, we have neglected some complex terms, such as $u^{*}v^2$ and $v^{*}u^2$. 
The parameter $c$ is assumed to be $c=1
/\{\sqrt{2}(1+g)\}$, because the symmetric solution $u=v$ propagates without changing the profile of the solitons.   
When $\eta$ is large, mutual interaction is weak and the two solitons propagate independently. However, as $y$ is close to the junction point $y=40$, the 
Josephson effect becomes strong because of a small $\eta$ value.  

\begin{figure}[tbp]
\begin{center}
\includegraphics[height=4.cm]{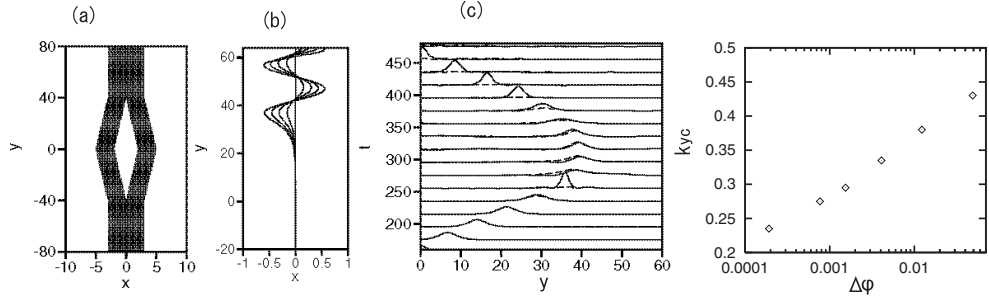}
\end{center}
\caption{(a) Channel with branching and junction points. (b) Trajectories of the center of gravity of solitons for $k_y=1$. The phase shifts of $\Delta \varphi=0, \pi/64,\pi/32$ and $\pi/16$ are given at $y=0$. The straight line is the trajectory for $\Delta \varphi=0$. The zigzag oscillation grows with $\Delta\varphi$. 
The trajectory that exhibits the largest zigzag oscillation corresponds to $\Delta\varphi=\pi/16$. 
(c) Time evolution of $|\phi|$ along the center lines of left and right channels at $k_y=0.37$ for $\Delta\varphi=\pi/256$. (d) Critical values $k_{yc}$ of the reflection as a function of $\Delta\varphi$.}
\label{f1}\end{figure}
\begin{figure}[tbp]
\begin{center}
\includegraphics[height=3.5cm]{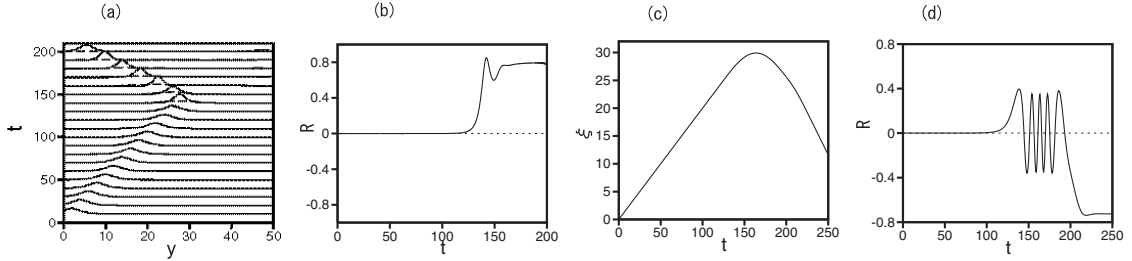}
\end{center}
\caption{(a) Time evolution of $|u|$ (solid curve) and $|v|$ (dashed curve) for $k_y=0.29$ and $\Delta \varphi=\pi/256$. (b) Time evolution of $R=(N_u-N_v)/(N_u+N_v)$. (c) Time evolution of $\xi(t)$ determined using  eq.~(7)  for $k_y=0.29$ and $\Delta \theta=\pi/128$. (d) Time evolution of $R$ determined using eq.~(7).}
\label{f2}
\end{figure}

Figure 2(a) shows the time evolution of $|u|$ and $|v|$ determined using the coupled equations (2) at $k_y=0.2$. The initial conditions of $u$ and $v$ are $u=A/{\rm cosh}(y/W)\exp(i\Delta\varphi)$ and $v=A/{\rm cosh}(y/W)\exp(-i\Delta\varphi)$ where $b=0.9,N_0=5/(\sqrt{\pi}b),W=4\sqrt{2}/N_0,A=\sqrt{N_0/(4W)}$, and $\Delta\varphi=\pi/256$. 
The two solitons are reflected, and one of the solitons dominates after the reflection. The critical velocity $k_y=0.295$ is slightly different from that in the direct numerical simulation, but the qualitative behaviors are similar. The asymmetry of the two solitons is expressed as $R=(N_u-N_v)/(N_u+N_v)$ where $N_u=\int |u|^2dy$ and $N_v=\int |v|^2dy$ are the norms of $u$ and $v$, respectively. The time evolution of the ratio $R$ is shown in Fig.~2(b). The ratio $R$ increases from 0 and  reaches a positive constant value after the reflection. This implies that one soliton dominates in the $u$-channel after the reflection. 

If $u(y,t)=A{\rm sech}\{(y-\xi)/W\}\exp\{ip(y-\xi)-i\theta_1\}$ and $v(y,t)=B{\rm sech}\{(y-\xi)/W\}\exp\{ip(y-\xi)-i\theta_2\}$ are further assumed, the effective Lagrangian is evaluated as
\begin{eqnarray}
L_{eff}&=&\frac{1}{2}\int \{ i(u_tu^*-uu_t^*+v_tv^*-vv_t^*)-|u_y|^2-|v_y|^2+c(|u|^4+|v|^4+2g|u|^2|v|^2)\nonumber\\& &+2d(uv^*+vu^*)-2d(|u|^2+|v|^2)\}dy\nonumber\\
&=&N_0[p\xi_t-p^2/2-1/(6W^2)+(\theta_{1t}+\theta_{2t})/2+R(\theta_{1t}-\theta_{2t})/2\nonumber\\& &+R^2N_0e/(16W^2)+\tilde{d}\{\sqrt{1-R^2}\cos(\theta_2-\theta_1)-1\}]
\end{eqnarray}
where $N_0=\int\{|u|^2+|v|^2\}dy=2(A^2+B^2)W$, $R=(A^2-B^2)/(A^2+B^2)$, $e(\xi)=\int c(y)\{1-g(y)\} {\rm sech}^4\{(y-\xi)/W\}dy$, and $\tilde{d}(\xi)=\int d(y){\rm sech}^2\{(y-\xi)/W\}dy/\int{\rm sech}^2\{(y-\xi)/W\}dy$.
The variational principle yields
\begin{eqnarray}
W&=&\frac{4}{\{1+R^2+(1-R^2)g\}N_0c},\nonumber\\
\frac{d\xi}{dt}&=&p,\nonumber\\
\frac{dp}{dt}&=&\frac{R^2N_0}{16W^2}\frac{\partial e}{\partial \xi}+\{\sqrt{1-R^2}\cos(\theta_2-\theta_1)-1\}\frac{\partial \tilde{d}}{\partial \xi},\nonumber\\
\frac{dR}{dt}&=&2\tilde{d}\sqrt{1-R^2}\sin\Delta\theta,\nonumber\\
\frac{d\Delta \theta}{dt}&=&\frac{RN_0e}{4W^2}-\frac{2\tilde{d}R}{\sqrt{1-R^2}}\cos\Delta\theta,
\end{eqnarray}
where $\Delta \theta=\theta_2-\theta_1$, and $\xi$ is the $y$-coordinate of the center of gravity of the solitons. 
If $R=0$ and $\theta_1=\theta_2$, $dp/dt=0$ and the two solitons propagate with a constant velocity. If $R<<1$ and $\Delta\theta<<1$, $d^2\Delta\theta/dt^2=-2\tilde{d}\{2\tilde{d}-N_0e/(4W^2)\}\Delta\theta$ is obtained using the last two equations in eq.~(4), which describes the Josephson oscillation.  If $N_0e/(4W^2)-2\tilde{d}>0$, the Josephson oscillation is amplified and the symmetric state $u=v$ becomes unstable. Because $e(\xi)$ and $\tilde{d(\xi})$ are not uniform, $dp/dt$ can become negative owing to the amplified Josephson oscillation. Figures 2(c) and 2(d) show the time evolution of $\xi$ and $R$ for $N_0=5/(\sqrt{\pi}b)$ with $b=0.9$ and $\xi(0)=0,p(0)=k_y=0.2$, and the initial phase difference $\Delta \theta(0)=2\Delta \varphi=\pi/128$. The reflection of the trajectory and the amplification of the Josephson oscillation are observed. The ratio $R$ takes a constant value  for large $t$ values, although the sign of the constant value depends strongly on the initial velocity $p(0)$ because of the many oscillations of $R$ near the reflection point.  The critical value $k_{yc}$ for $\Delta\theta(0)=\pi/128$ determined using eq.~(4) is 0.305, which is close to that by eq.~(1). The variational approximation is fairly good.  

To summarize, we have performed some numerical simulations of channel-guided solitons in the two-dimensional nonlinear Schr\"odinger equation. We have found that reflection occurs near the junction induced by the amplification of the Josephson oscillation.  The complex dynamical behaviors can be analyzed approximately on the basis of the variational principle using effective Lagrangians. Channel-guided solitons exhibit various complex dynamics, and we would like to investigate this further in the future.   

We would like to thank Prof.~B.~A.~Malomed for valuable discussions. 


\begin{thebibliography}{99}
\bibitem{rf:1} L.~F.~Mollenauer and J.~P.~Gordon: {\it Solitons in Optical Fibers} (Academic Press, San Diego, 2006).
\bibitem{rf:2} J.~Denschlag, J.~E.~Simsarian, D.~L.~Feder, C.~W.~Clark, L.~A.~Collins, J.~Cubizolles, L.~Deng, E.~W.~Hagley, K.~Helmerson, W.P.~Reinharts, S.~L.~Rolston, B.~I.~Schneider, and W.~D.~Phillips: Science {\bf 287} (2000) 97. 
\bibitem{rf:3} L.~Kaykovich, F.~Schreck, G.~Ferrari, T.~Bourdel, J.~Cubizolles, L.~D.~Carr, Y.~Castin, and C.~Salomon: Science {\bf 396} (2002) 1290. 
\bibitem{rf:4} H.~Sakaguchi and B.~A.~Malomed: Phys. Rev. A {\bf 75}  (2007) 063825.
\bibitem{rf:5} H.~Sakaguchi and Y.~Kageyama: J. Phys. Soc. Jpn. {\bf 79} (2010)  113002.
\bibitem{rf:6} D.~Cassettari, B.~Hessmo, R.~Folman, T.~Maier, and J.~Schmiedmayer: Phys. Rev. Lett. {\bf 85} (2000) 5483. 
\bibitem{rf:7} M.~W.~Jack, M.~J.~Collett, and D.~F.~Walls: Phys. Rev. A 
{\bf 54} (1996) R4625.
\bibitem{rf:8} H.~Sakaguchi and B.~A.~Malomed: Phys. Rev. A {\bf 83} (2011) 036608. 
\end{thebibliography}
\end{document}